\documentclass[letterpaper,twocolumn,10pt]{sig-alternate}
\usepackage{subfigure}                  
\usepackage{multirow}                   
\usepackage{enumitem}                   
\usepackage{balance}                    

\usepackage{amsmath}

\usepackage{filecontents}

\usepackage{amssymb}
\usepackage{inconsolata}
\usepackage{array}
\usepackage{booktabs}
\usepackage{fullpage}
\usepackage{threeparttable}
\usepackage{wasysym}
\usepackage{flushend}
\usepackage[font={small}]{caption}
\usepackage{longtable}
\usepackage{scrtime}
\usepackage[multiple]{footmisc}
\usepackage{footnote}
\usepackage{makecell}
\usepackage{multirow}
\usepackage{rotating}
\usepackage{multicol}
\usepackage{graphicx}
\usepackage{float}
\usepackage{listings}
\usepackage{textcomp}
\usepackage{enumitem}
\usepackage{tikz}
\usepackage{verbatim}
\usepackage{booktabs}
\usepackage{siunitx}
\usepackage{pifont}
\usepackage{microtype}
\usepackage[compact]{titlesec}
\usepackage[toc,page]{appendix}
\usepackage{balance}
\usepackage{adjustbox}
\usepackage{xcolor}
\usepackage{soul}
\usepackage{epsfig,hyperref,color,graphicx,xspace,pifont,dirtree}
\usepackage{cleveref}

\crefformat{section}{\S#2#1#3} 
\crefformat{subsection}{\S#2#1#3}
\crefformat{subsubsection}{\S#2#1#3}

\hypersetup{%
  linkcolor=black
}

\usepackage{fancyhdr}
\usetikzlibrary{shapes.geometric, arrows}
\usepackage{pgfplotstable}

\usepackage{pgfplots}
    \pgfplotsset{compat=1.3}
    
    \definecolor{bblue}{HTML}{4F81BD}
    \definecolor{rred}{HTML}{C0504D}
    \definecolor{ggreen}{HTML}{9BBB59}

\newcommand{\rom}[1]{{\em\lowercase\expandafter{(\romannumeral #1\relax)}}}
\newcommand{\nom}[1]{{\em\lowercase\expandafter{(#1\relax)}}}

\definecolor {processblue}{cmyk}{0.80,0,0.20,0.30}
\definecolor{darkgreen}{rgb}{0.18,0.54,0.34}
\definecolor{maroon}{rgb}{0.64,0.16,0.16}
\definecolor{darkpink}{rgb}{0.75,0.25,0.5}
\definecolor{backcolour}{rgb}{0.90,0.95,0.94}

\definecolor{b1}{RGB}{31,119,180}
\definecolor{b2}{RGB}{23,190,255}
\definecolor{tbrown}{RGB}{140,86,75}
\definecolor{tpink}{RGB}{227,119,194}
\definecolor{tgreen}{RGB}{44,160,44}
\definecolor{tpur}{RGB}{148,103,189}
\definecolor{tyel}{RGB}{188,189,34}
\definecolor{tgray}{RGB}{127,127,127}
\definecolor{tred}{RGB}{214,39,40}
\definecolor{torange}{RGB}{255,127,14}

\lstdefinestyle{customc}{
    backgroundcolor=\color{white},   
   breaklines=true,
  language=C,
numbers=left,  
xleftmargin=2.2em,
frame=single,
framexleftmargin=2em,
  basicstyle=\ttfamily\scriptsize,
  showstringspaces=false,
  keywordstyle=\color{blue}\bfseries,,
  identifierstyle=\color{black}\bfseries,,
  stringstyle=\color{blue}\bfseries,,
  commentstyle=\itshape\color{darkgreen}\bfseries,,
    tabsize=2,
    morekeywords={include, typeof, static, new, begin, end, struct, char, void, unsigned, long, const, privilege_enclave, int, true, false, catch, function, return, null, catch, switch, var, if, in, while, do, else, case, break}
}
\newcommand{\tool} {{Argus}\xspace}

\newcommand{\kmod} {{TPMS}\xspace}

\newcommand{\squeezeup}{\vspace{-5mm}}

\definecolor{lightgray}{rgb}{.95,.95,.95}
\definecolor{darkgray}{rgb}{.4,.4,.4}
\definecolor{purple}{rgb}{0.65, 0.12, 0.82}

\graphicspath{{./figure/}}

\begin{document}

\title{Enhancing the Security \& Privacy of Wearable Brain-Computer Interfaces}

    {}


\author{
{Zahra Tarkhani}\thanks{\hphantom{} Equal contribution.}\\
University of Cambridge
\and
{Lorena Qendro$^*$}\\
University of Cambridge
\and
{Malachy O'Connor Brown}\\
University of Cambridge
\and
{Oscar Hill}\\
University of Cambridge
\and
{Cecilia Mascolo}\\
University of Cambridge
\and
{Anil Madhavapeddy}\\
University of Cambridge
} 

\maketitle

\begin{abstract}
Brain computing interfaces (BCI) are used in a plethora of safety/privacy-critical applications, ranging from healthcare to smart communication and control.
Wearable BCI setups typically involve a head-mounted sensor connected to a mobile device, combined with ML-based data processing. Consequently, they are susceptible to a multiplicity of attacks across the hardware, software, and networking stacks used that can leak users' brainwave data or at worst relinquish control of BCI-assisted devices to remote attackers.

In this paper, we:
\emph{(i)} analyse the whole-system security and privacy threats to existing wearable BCI products from an operating system and adversarial machine learning perspective; and
\emph{(ii)} introduce \tool, the first information flow control system for wearable BCI applications that mitigates these attacks.
\tool' domain-specific design leads to a lightweight implementation on Linux ARM platforms suitable for existing BCI use-cases. Our proof of concept attacks on real-world BCI devices (Muse, NeuroSky, and OpenBCI) led us to discover more than $300$ vulnerabilities across the stacks of \emph{six} major attack vectors. Our evaluation shows \tool is highly effective in tracking sensitive dataflows and restricting these attacks with an acceptable memory and performance overhead ($<15\%$).

\end{abstract}

\section{Introduction}
\label{sec:intro}

A rapidly expanding set of mobile brain-computer interfaces (BCI) deliver solutions for monitoring health, mental/emotional state (e.g., focus, anxiety, or motivation), or sleep quality~\cite{lakhan2019consumer,abdulkader2015brain}.
Lately, relatively low-cost wearable BCIs are customized to support people with physical impairments or to improve smart living through BCI-assisted drones, robotic arms, wheelchairs, or mixed reality environments~\cite{fatima2015towards,kim2014quadcopter,jafri2019wireless}.
In addition, several proposals for utilizing BCI for multi-factor authentication and crypto-biometrics~\cite{pham2014multi,jayarathne2016brainid,landau2020mind} are emerging.
Compromising BCI-controlled devices---such as by taking control of a wheelchair or drone---is a real physical threat. Also, sensitive personal information, like thoughts, sexual orientation, or religious beliefs, is under threat if security and privacy measures are not fully adopted~\cite{bernal2021security}. 

\begin{figure}[t]
\centering

\vspace{0.25in}
\includegraphics[scale=.5]{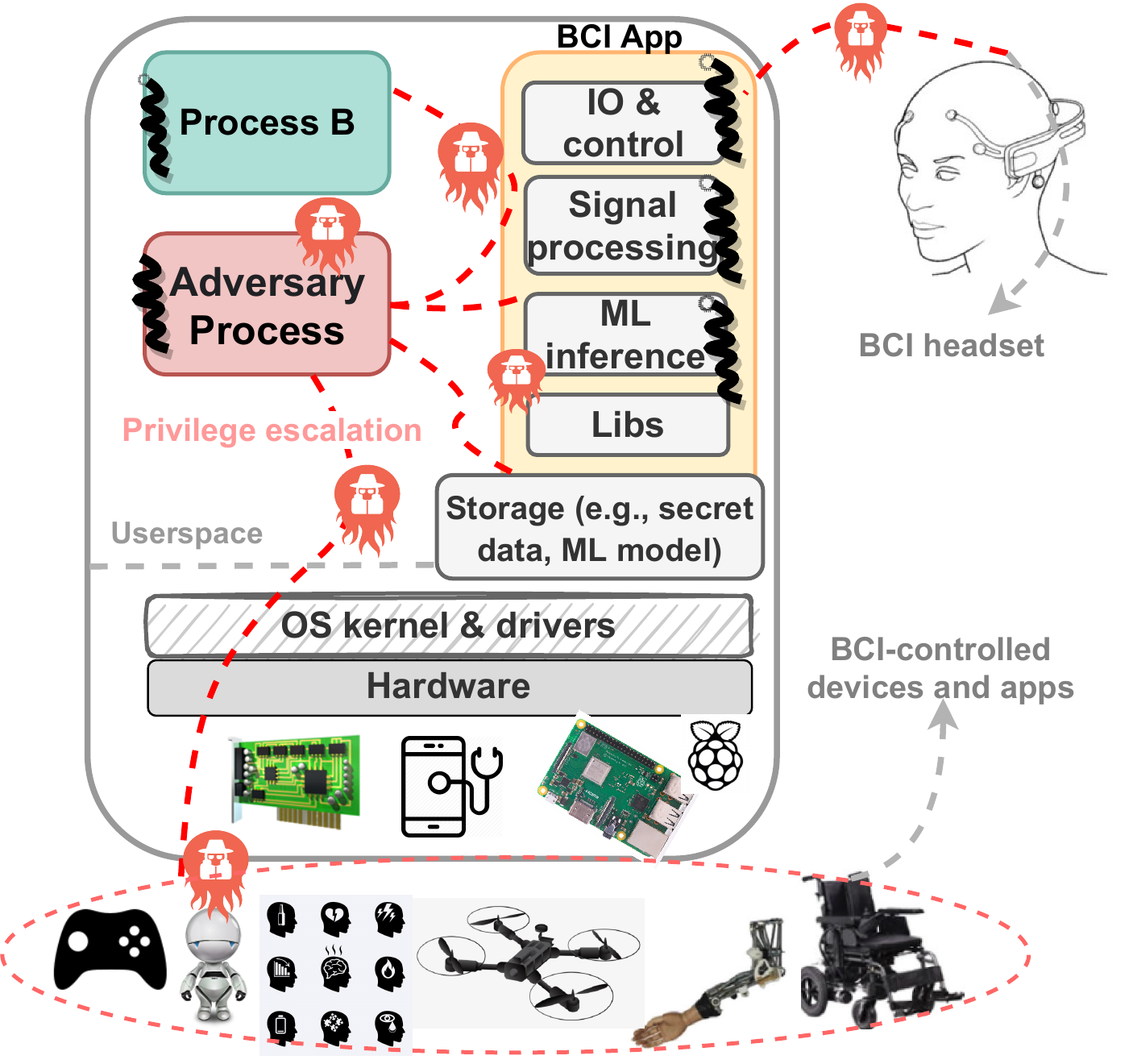}
\caption{\small{Attack surface of BCI-edge platforms.}}
\label{fig:attsurface}
\squeezeup
\end{figure}

Figure~\ref{fig:attsurface} illustrates how attackers can exploit and compromise BCI applications at different stages, and consequently steal secrets or take control of the host or BCI-controlled devices.
Unauthorized access to pre-trained deep learning models on the host device could lead to the injection of malicious inputs, or the leaking of prediction model parameters or training data~\cite{shokri2017membership, mo2020darknetz, qendro2021stochastic}. Attackers can also combine/use traditional system-based attacks with/for ML adversarial ones to gain private information or manipulate ML predictions. Simple manipulation of BCI peripherals can alter the ML model output (e.g., from drowsy to awake) causing serious traffic accidents in EEG-based driving assistance for self-driving cars~\cite{zhu2021vehicle}.

Previous studies discuss the possibility and consequences of attacking BCI applications~\cite{bernal2021security,bonaci2014app, meng2019white,zhang2021tiny}. However, their considered attack vectors are limited, and very few investigate real-world BCI use cases. Additionally, these works do not offer any solution. In this paper, we overcome the limitations of previous work by presenting prototypes for six key attack vectors, including the combination of the system-side and ML attacks, on real-world BCI applications.
We observe that besides system-side attacks, currently feasible adversarial attacks on BCI-edge environments could also be detected by carefully modeling and mapping the attack vectors into the underlying system objects (e.g., files, memory regions, device peripherals) and then monitoring the objects. 
Hence, we propose a principal approach for 
enhancing security and privacy on BCI-edge platforms. 
To achieve this, we introduce \tool, a \textit{domain-specific} information flow tracking system, for reducing both system-based and ML adversarial attack surface in a lightweight manner.

Previous information flow tracking systems (such as TaintDroid~\cite{enck2014taintdroid}, Weir~\cite{badger1995practical}, or FlowFence~\cite{fernandes2016flowfence}) improve security and privacy on edge platforms. However, these solutions come with a significant complexity and overhead, particularly when dealing with non-trivial security policies (e.g., in-process threats) and handling large number of tainted objects. They are either strictly designed for Android or force the adoption of a specific programming language. 
Unlike these systems, \tool is designed to trace fine-grained system objects (such as memory regions, threads, files) inside different parts of a BCI applications without depending on a specific programming language.
Enabling this through a novel OS kernel module and userspace framework allows us to achieve efficiency, adding just a small performance overhead and memory footprint.

Our major contributions are as follows:

\begin{itemize}[leftmargin=*]
    \vspace{-0.08in}
    \item Comprehensive security and privacy investigation of contemporary wearable BCIs. This includes finding more than 300 vulnerabilities and presenting real-world PoCs for six key attack vectors (\S\ref{sec:state}). 
    \vspace{-0.18in}
    \item Introducing \tool, a novel domain-specific framework for tracking sensitive information flows in modern BCI-edge use cases. \tool is the first framework to support both system and adversarial ML attacks (\S\ref{sec:over}).
    \vspace{-0.08in}
    \item Efficient implementation of \tool on Linux-ARM platforms. This is achieved through a kernel module for tracing fine-grained system objects, and a userspace library (Lib\tool) for specifying complex security policies (\S\ref{sec:implementation}).
    \vspace{-0.08in}
    \item Evaluating \tool with microbenchmarks and integration to real-world wearable BCIs. Our results demonstrate \tool' effectiveness on reducing different attack vectors while adding a reasonable performance and memory overhead (\S\ref{sec:eval}). 

\end{itemize}

\section{Related Work}
\label{sec:related}

Our work is inspired by the lack of a thorough investigation of attacks towards wearable BCIs and the limitations of existing techniques at detecting and mitigating them. Here we summarise the differences between our work and existing systems.


\textbf{Security \& privacy threats on weareble BCI.} While former studies discuss the possibility and consequences of attacking BCI applications~\cite{bernal2021security,bonaci2014app}, very few investigate real-world BCI use cases.
Brain spyware~\cite{martinovic2012feasibility} presents a malicious software to communicate visual stimuli and steal private information (e.g., 4-digit PINs, or location of residence) using a BCI to record the EEG signals. Takabi et al.~\cite{takabi2016brain} highlighted inadequate privacy considerations in the NeuroSky and Emotive APIs. More recently, other research efforts explored adversarial attacks on EEG-based models~\cite{meng2019white,zhang2021tiny, qendro2021early}. 
However, unlike this work, these studies limit their investigation to only a few simple attack vectors without considering the combination of system and ML adversarial attacks. Additionally, they do not offer any solution for effectively enhancing the security and privacy of wearable BCI.

\textbf{Information flow tracking.} There are various techniques for controlling dataflows
to improve security and privacy ~\cite{myers1997decentralized,wang2015between, krohn2007information,zeldovich2008securing,egele2011pios,pasquier2015camflow,egele2011pios,jiang2013detecting,tarkhani2020enclave}. Particularly for mobile/edge systems, TaintDroid~\cite{enck2014taintdroid} shows the effectiveness of dataflow tracking of privacy-sensitive data in Android applications. Similarly, Weir~\cite{nadkarni2016practical} extended Android with decentrised IFC-based access control, while FlowFence~\cite{fernandes2016flowfence} presents an IFC-based data protection system for IoT. While these solutions inspire our work, they demonstrate a significant complexity and overhead of general-purpose IFC, particularly when handling a large number of tainted objects and non-trivial security policies. 
Additionally, language-level IFC, such as TaintDroid, forces the use of a specific programming paradigm (e.g., labelling Java classes or objects).
\tool is optimized to track information flows for wearable BCI use cases with small overhead and no specific language dependency. Moreover, unlike general-purpose applications, BCI apps require a considerably smaller number of tainted objects, specially in-application memory regions. This leads to tracking in-process system objects more efficiently using hardware-based optimizations (\S\ref{sec:implementation}).

\textbf{Attack investigation frameworks.} There exist systems for detecting and exploring security threats through services such as debugging, fuzzing auditing, province tracking, and logging systems~\cite{ji2017rain, ma2016protracer,suhail2016introducing,xia2015effective,brasser2016regulating,mirzamohammadi2017ditio,bates2015trustworthy,muniswamy2006provenance,pohly2012hi}. However, this works' goals and \tool' design principles differ. This paper focuses on two key points: (i) firstly, presenting real-world attack PoCs based on adversarial ML and system/app adversaries (and their combinations) for wearable BCI use cases; and (ii) secondly, a novel information flow tracking system is introduced to mitigate these attack vectors.

\textbf{Domain-specific security and privacy extensions.}
The current state of mobile/IoT security threats is too big for a general-purpose solution. As a result, recently, we have observed a surge in domain-specific solutions for improving security and privacy on mobile/IoT use cases~\cite{talebi2021megamind,lei2020secwir,beck2020privaros,harris2018aggio,fernandes2016security}. To name a few, Privaros~\cite{beck2020privaros} enforces host-specified privacy policies on delivery drones, while McReynolds et al. tackle privacy issues on children's smart toys~\cite{mcreynolds2017toys}, and Aggio \cite{harris2018aggio} focuses on privacy-preserving smart retail environments. Additionally, MegaMind~\cite{talebi2021megamind} offers a solution targeted to voice assistants, SpecEye~\cite{li2019speceye}  proposes a privacy-preserving screen exposure detection system, and SecureSIM~\cite{zhao2021securesim} provides an access control specific to in-SIM files.
Like these systems, we believe it is essential to investigate and provide solutions for specific highly sensitive domains, wearable BCI in our case.

\section{Security \& privacy threats on wearable BCIs}\label{sec:state}


Here, we summarise our findings and attack prototypes on real-world BCI applications.
\subsection{Overview of BCI platforms}

BCI platforms consist of four primary units: \nom{1} brain signal acquisition, \nom{2} feature extraction and signal processing, \nom{3} data analytics and ML computation, and \nom{4} output prediction, user feedback, or control signals. 
In this paper, we focus on non-invasive BCI techniques based on electroencephalography (EEG) data\footnote{There are solutions combining EEG with fMRI, and fNIRS~\cite{alzahab2021hybrid}. However,
EEG is portable and relatively inexpensive (especially compared to fMRI).} that form most BCI applications. Additionally, we use commodity devices such as Muse~\cite{muse}, NeuroSky~\cite{neurosky}, and OpenBCI~\cite{openbci}.

As Figure~\ref{fig:attsurface} shows, raw EEG data is transferred from the BCI headset (e.g., via Bluetooth) and processed through an application running on a smartphone or microprocessor like Raspberry Pi. Modern BCI applications rely on predictive algorithms ranging from simple ML linear models to recent pre-trained deep learning models for better inference~\cite{hosseini2017optimized,lv2020advanced,zgallai2019deep,craik2019deep}.
These applications also depend on third-party libraries, SDKs, and OS (mostly Linux-based) services to communicate with the BCI headsets and to control external devices. Here, we describe six attack vectors on various layers of computation and interactions with the host.

\begin{table}[t]
\caption{Summary of vulnerabilities we found in BrainFlow.}
\resizebox{\columnwidth}{!}{%
  \footnotesize
    \begin{tabular}{|l|l|l|}
    \hline
    Vulnerability type & Numbers & Description                                \\ \hline
    CWE-367            & 11      & Time-of-check Time-of-use (TOCTOU)         \\
    CWE-134            & 34      & Use of externally-controlled format string \\
    CWE-120/CWE-119    & 317     & Memory Corruption/buffer overflow          \\
    CWE-126            & 38      & Buffer over-read                           \\
    CWE-20             & 16      & Improper input validation                  \\
    CWE-362            & 33      & Improper synchronisation/race condition    \\ \hline
    \end{tabular}}
    \label{tab:bflow}
\end{table}

\subsection{Threats from/to the host}

\textbf{(AV1) Sniffing, Spoofing \& Man-in-the-middle.}
Due to unencrypted and insecure BLE (Bluetooth Low Energy) connections with the BCI host devices, numerous sniffing attacks are possible. For example, we implemented several sniffers in all the considered BCI devices to capture and record all transmitted packets, MAC addresses, and connection parameters. Gaining this information facilitates the launch of more complex attacks such as MITM. Via the MITM attack, we could easily intercept and record all data sent between the headset and device and, additionally, compromise the device integrity by altering the data. As we explain in \S\ref{sec:hostatt}, one of our prototypes hijacks and alters communications between the headset and phone using an unauthorized Raspberry Pi acting as the BLE peripheral device. 

\textbf{(AV2) Inadequate isolation \& access control.}
All BCI applications and devices we investigated lack adequate access control mechanisms. By launching an adversary process, we could easily access and modify the application stored data and deep learning model on the file system. Then we could leak secrets (e.g., predicted mental state of the user) to the outside, alternating control peripherals/signals (e.g., to change the direction of a BCI-controlled drone), or maliciously target the deep learning model by sending crafted queries~\cite{shokri2017membership, sanchez2020game}. Similarly, we could compromise the applications through insecure interactions with the host, for example, via inter-process communication or shared memory.

\textbf{(AV3) Privilege escalation.}
Wearable BCI platforms suffer from poor privilege management. They do not use any isolation mechanism to protect their own resources from other applications, neither separate privileges within their different components (e.g., by isolating untrusted libraries). Moreover, some require higher privilege such as ``root-access'' for their operations (e.g., OpenBCI's user interface), that can be misused by attackers to take control of the host. We implemented several privilege escalation attacks to leak information and successfully launched confused deputy attacks to compromise other host applications and even the host OS through an insecure BCI application (\S\ref{sec:hostatt}). Additionally, to show the importance of in-application compartmentalisation, we analysed third-party libraries such as BrainFlow~\cite{brainflow} (see Table~\ref{tab:bflow}).
As expected, by exploiting the vulnerabilities within untrusted dependencies, we could gain unauthorised privileges.

\subsection{ML adversarial threats}
\label{sec:adversarial_ml}

\begin{figure}[t!]
    \centering
    \includegraphics[width=0.5\textwidth]{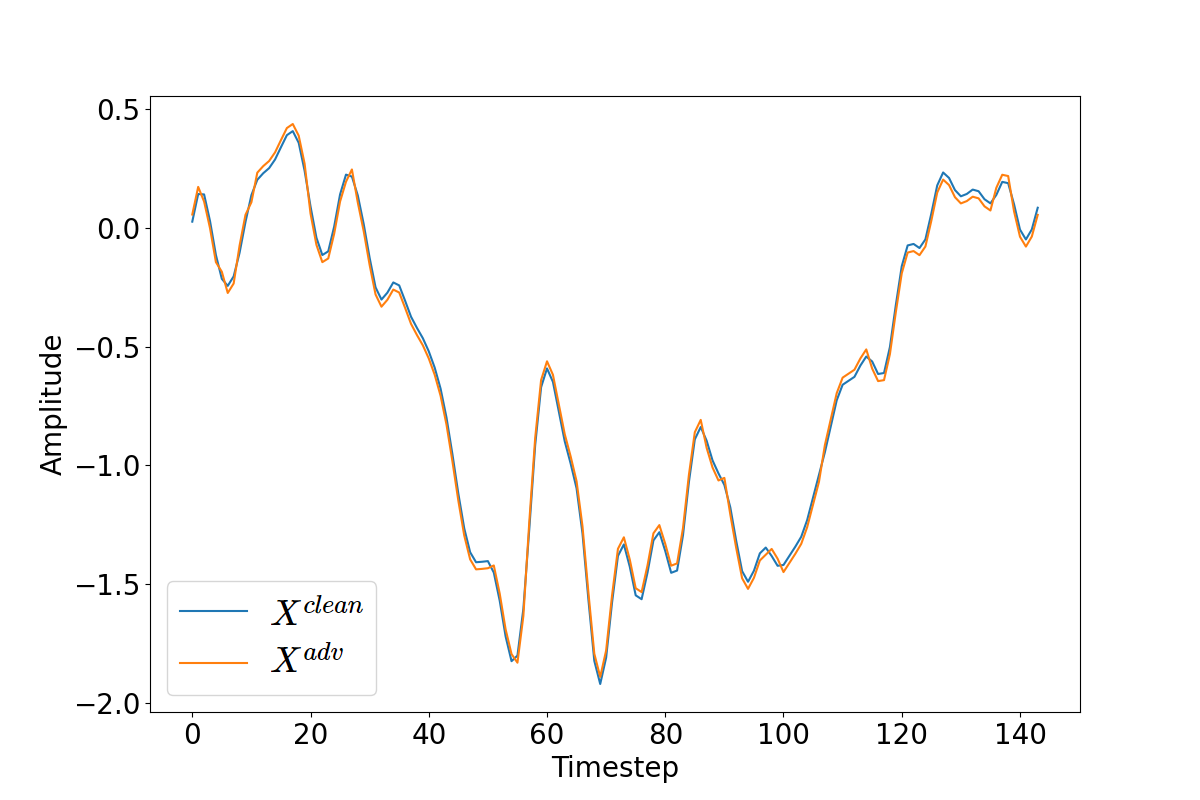}
    \vspace{-0.35in}
    \caption{PGD attack (40 iterations, $\epsilon=0.5$) on an EEG sample of the P300 dataset. Perturbed signal is close to the physiologically plausible signal but the final prediction differs.}
    \label{fig:clean_adv}
\end{figure}

\textbf{Adversarial attacks overview.} Current wearable BCI devices are at early stages of adopting deep learning algorithms. They are, however, rapidly following the trend of enabling AI on mobile/wearable environments (e.g., smart watches). Therefore, we aim at providing a framework that supports investigating adversarial deep learning attack vectors, too.
Deep learning models can be manipulated since current implementations overlook robustness towards adversarial attacks. To prevent this issue, our framework also targets adversarial attacks widely used against EEG-based deep learning models. Here, we focus on inference-time attacks given the computational capacity of BCI systems (i.e. unable to handle training on device). As such, these attack vectors force the model to misclassify the input sample. The most common inference-time attacks are black-box or white-box attacks, depending on the access level they have on the model or training data.

For white-box attacks, the attacker knows everything about the target model (e.g., architecture, training data and parameters). These attacks are more expensive to implement, however, mitigation techniques towards them are expensive and, often, unreliable too. 
Black-box attacks represent random or uninformed perturbations to the signal by having access only to the network input and output. They are often implemented by exploiting vulnerabilities shared between different models. We prototyped two EEG-based state-of-the-art white-box attacks such as \emph{Fast Gradient Sign Method} and \emph{Projected Gradient Descent} (widely used in computer vision) as well as introduced a PoC for a new EEG signal specific black-box attack which we call \emph{peripheral injection attack}.

\begin{figure}[t!]
    \centering
    \includegraphics[width=0.5\textwidth]{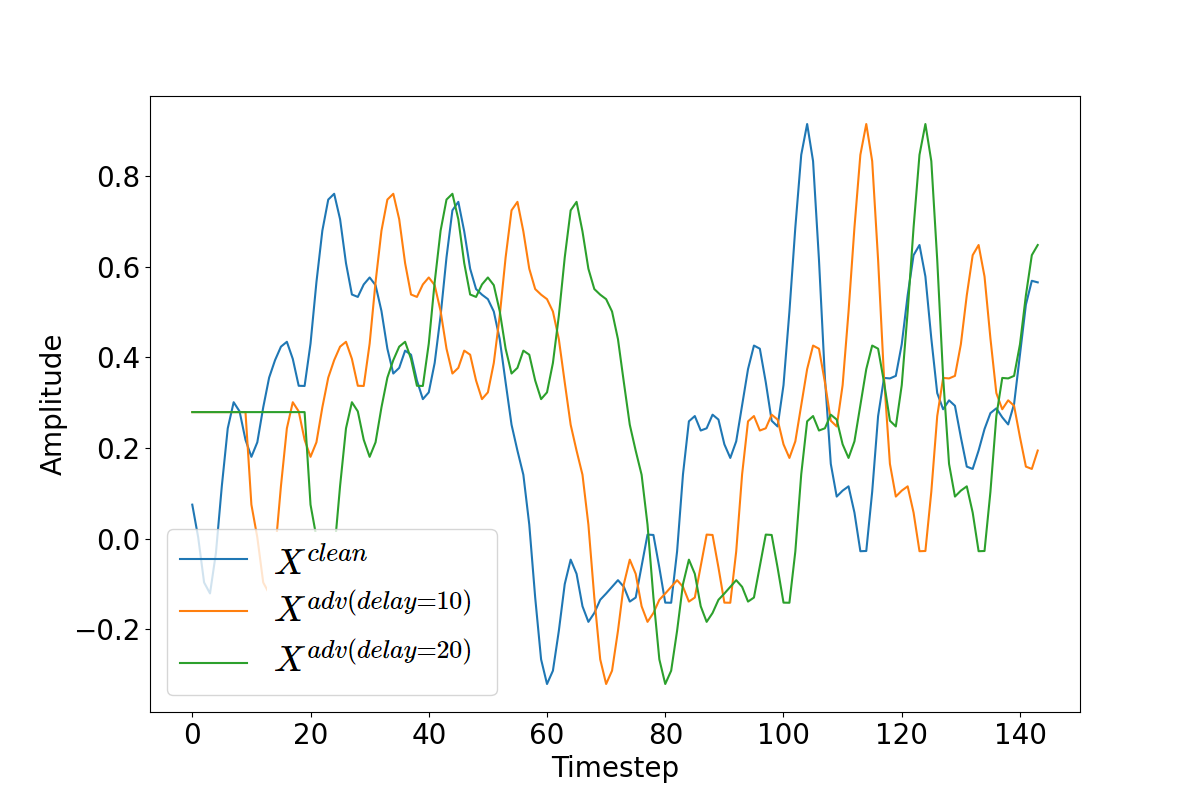}
        \vspace{-0.35in}
    \caption{Peripheral injection attack on an EEG sample of the P300 dataset. Slight delay introduced at the signal start can decrease accuracy.}
    \label{fig:peripheral_adv}
\end{figure}

\begin{figure*}[t]
\centering
\includegraphics[width=.98\textwidth]{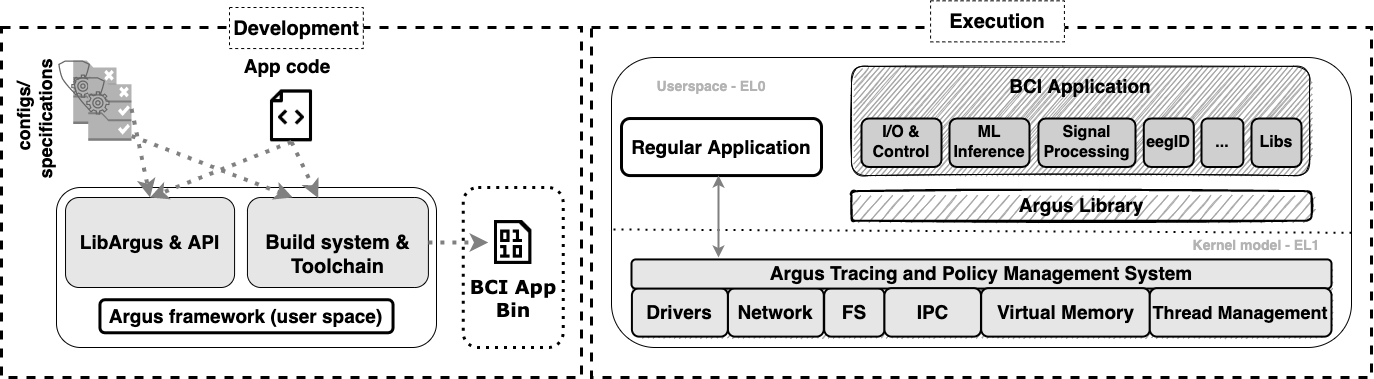}
\caption{\small{\tool high-level architecture.}}
\label{fig:argus_overview}
\squeezeup
\end{figure*}

\textbf{(AV4) Fast Gradient Sign Method (FGSM). }
Given a clean signal $\mathbf{X}$, an adversarial attack introduces small perturbations $\nabla$ such that the prediction for $\mathbf{X}$ and $\mathbf{X}^{adv}$ differs. FGSM~\cite{goodfellow2014explaining} is a single step attack which aims to find the adversarial perturbations by moving in the opposite direction to the gradient of the loss function $L(\mathbf{X}, y)$ w.r.t. the signal ($\nabla$):
\begin{equation*}
\small \mathbf{X}^{adv} = \mathbf{X} + \epsilon * sign(\nabla_{\mathbf{X}}L(\mathbf{X}, y)),
\end{equation*}
where $\epsilon$ is the step size which restricts the $l_{\infty}$ of the perturbation. \\

\textbf{(AV5) Projected Gradient Descent (PGD).} 
A stronger variant of FGSM~\cite{kurakin2016adversarial} consists of applying it iteratively introducing a small step $\alpha$:
\begin{equation}
\small
    \mathbf{X}^{adv}_0 = \mathbf{X'}, \mathbf{X}^{adv}_{n+1} = clip^{\epsilon}_{\mathbf{X}} \{\mathbf{X}^{adv}_{n} + \alpha sign\big(\nabla_{\mathbf{X}}L\big(\mathbf{X}^{adv}_{n}, y \big) \big)\}
    \label{bim}
\end{equation}
where
\begin{equation}
\small
    \mathbf{X'} = \mathbf{X} + \epsilon_1 * sign\big(\mathcal{N}\big(\mathbf{0}^d,\mathbf{I}^d\big)\big)
    \label{rfsgm}
\end{equation}

(with parameters $\epsilon_1$, $\epsilon$ such as $\epsilon_1 < \epsilon$) is an additional prepended random step~\cite{tramer2017ensemble} which avoids going towards a false direction of ascent. Steps 1 and 2 make PGD~\cite{madry2017towards}, a universal first-order attack.

We launched an adversary process to access the BCI application file system and get the deep learning model as well as the input data from the headset (see AV2). We further applied the two aforementioned attacks (at different instances) to create a perturbed signal which misclassifies the input to a different (targeted and non-targeted) class.  In \S~\ref{sec:hostatt}, we show that the white-box attacks (AV5 and AV6) can have adversarial success rate of 100\% and decrease the accuracy from 70\% to 0\% hindering the trust and deployability of the automatic deep learning system. Moreover, as shown in Figure~\ref{fig:clean_adv}, they can be very difficult to detect by (non-)expert observations as the perturbed signal is very close to the physiologically plausible signal.

\textbf{(AV6) Peripheral injection black-box attack}. Given the lack of security features in the studied BCI platforms, we launched an AV1 to hijack the headset-phone communication and created a black-box attack by introducing a signal delay as adversarial perturbation (without accessing the deep learning model). 
Detecting this category of attacks is not straightforward since 
delays in the signal can also incur naturally, therefore, we need a tracing system that can identify malicious behaviour.
Given an EEG timeseries $\mathbf{X}(t)$, a time-shift can delay or advance the signal in time by an interval $\pm\tau$ such as $\mathbf{X}^{adv}(t) = \mathbf{X}(t\pm\tau)$. We first calculate the mean of the signal along its final axes, then we shift the signal over $\tau$ steps, where elements shifted beyond the last position are re-introduced at the first position (see Figure~\ref{fig:peripheral_adv}). Finally, we set the perturbed signal to be:
\vspace{-0.05in}
\begin{equation*}
\small
    \mathbf{X}^{adv}(t) =
    \begin{cases}
    mean(\mathbf{X}) & t < \tau \\
    \mathbf{X}(t - \tau) & t \ge \tau
    \end{cases}    
    \vspace{-0.05in}
\end{equation*}
This black-box attack is fast and easy to implement, however, as we'll see in \S~\ref{sec:hostatt}, it can heavily impact the availability (e.g., to form a denial-of-service attack (DoS)) and reliability (e.g., alter class prediction or lower confidence) of the BCI deep learning model. 


\begin{table*}[t]
\caption{A subset of simplified policies used for IFC monitoring on our attack PoCs. We use subscripts $\otimes$ for attacker system objects, $b$ for BCI app, and $a$ for any host app. $\longrightarrow$ denotes a permitted information flow, while $\not\longrightarrow$ represents an information flow policy violation. Here $P\{\}$ represents a set of processes/threads, $F\{\}$ a set of files, $S\{\}$ a set of sockets (and their binding ports), $M\{\}$ a set of memory regions, $IPC\{\}$ a set of any object used for inter-process communication (e.g., pipes or shared memory). [For clarity, we omit the set annotation $\{\}$.]}

\centering
\resizebox{2\columnwidth}{!}{%
   \begin{tabular}{|l|l|l|l|l|}
    \hline
    \textbf{Attacks} & \textbf{Tainted objects ($O_b$)} & \textbf{Secrecy policy ($\forall obj \in O_b$}) & \textbf{Integrity policy ($\forall obj \in O_b$})  \\ \hline

AV1 & $F_b\{f_{in},f_{out},f_{dev}\}\cup S_b\{port_{in},port_{out}\} \cup IPC_b$ &  $obj\not\longrightarrow P_{\otimes} \land P_{a}$  & $P_{\otimes} \land P_{a} \not\longrightarrow  obj$                             \\ \hline

 AV2 & $P_b \cup F_b \cup S_b  \cup IPC_b  \cup M_b$ &  $obj \not\longrightarrow P_{\otimes} \land P_{a} $  & $P_{\otimes}\land P_{a} \not\longrightarrow obj$                             \\ \hline

AV3 & $P_b \cup F_b \cup S_b  \cup IPC_b  \cup M_b$ & if $\exists obj \in O_b: obj \longrightarrow P_a$ & 
     if $P_a \longrightarrow obj,\exists obj \in O_b $  \\ 
 &  & then $P_a \not\longrightarrow P_{\otimes} $ & 
     then $P_{\otimes} \not\longrightarrow P_a $  \\ \hline
                
AV4 & $P_b \cup F_b\{f_{dev},f_{in}\} \cup S_b\{port_{in}\}$ & $ obj\not\longrightarrow P_{\otimes} $  & $P_{\otimes} \not\longrightarrow obj$                             \\ \hline

AV5 & $P_b \cup F_b\{f_{model},f_{dev},f_{in}\} \cup S_b\{port_{in}\}$ & $obj\not\longrightarrow P_{\otimes} $  & $P_{\otimes} \not\longrightarrow obj$                             \\ \hline
    
AV6 & $P_b \cup F_b\{f_{dev}\} \cup S_b\{port_{in}\}$ & $obj\not\longrightarrow P_{\otimes} $   & $P_{\otimes} \not\longrightarrow obj$                             \\ \hline
    \end{tabular}

}
\label{tab:detect}
\vspace{-0.10in}
\end{table*}

\section{Designing \tool }
\label{sec:over}
Figure~\ref{fig:argus_overview} shows the architecture of \tool consisting of a userspace library (Lib\tool) and an OS kernel module for information flow control and policy management for BCI applications. 

Lib\tool exposes the \tool API enabling tracing policies on a variety of system objects. Lib\tool is customizable, since it provides low-level APIs (e.g., \texttt{pthread, malloc,} or \texttt{fopen} granularity) for a modular investigation of different attack vectors.

To ensure better security and efficiency the \tool tracing and policy management system (\kmod) runs inside the OS kernel. Its role is to, firstly, capture the high-level specifications from Lib\tool, converting them into associated policies which are further enforced
based on our underlying \textit{unified} tracing model (\S\ref{sec:tmodel}). \kmod is designed to support customizable policies and tracking granularities, ranging from fine-grained in-application modules to application-wide policies. 
Secondly, it supports information flow tracking over a rich set of system objects, including address spaces, threads, files, sockets, and pipes without human intervention at runtime. 
BCI applications are characterised by a limited number of modules and dependencies which is ideal to provide an optimization for tracking fine-grained system objects (e.g., memory domains) in an efficient manner (see \S\ref{sec:implementation}). Since, \kmod creates the building blocks for tracing and monitoring arbitrary system objects, it is extensible and customizable to include new objects or remove existing ones. 

\begin{figure}[t]
\centering
\includegraphics[width=.9\columnwidth]{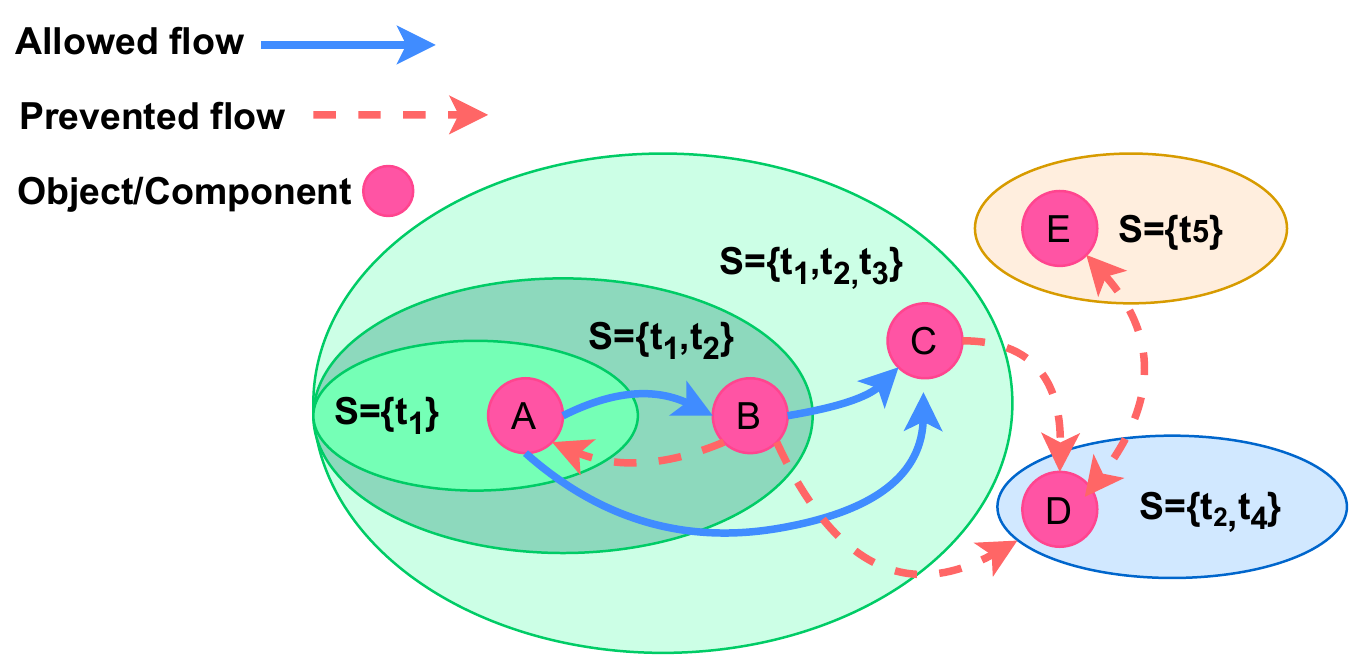}
\vspace{-0.10in}
\caption{\small{\tool: checking secrecy information flows.}}
\label{fig:sflow}
\squeezeup
\end{figure}
\subsection{Design principles}\label{sec:goals}
\tool aims to achieve the following design principles:

     \textbf{Extensibility.} Targeting different classes of attacks, particularly those combining different attack vectors (e.g., ML adversarial through bypassing BCI peripheral access control), requires supporting a complex set of security policies. Hence, \tool should not rely only on a fixed model and limited set of policies, instead it should allow for extending and customizing them depending on the threat model. 

    \textbf{Fine-granularity.} It is essential to track system objects over different parts of a BCI application (e.g., to detect malicious third-party libraries). This requires supporting fine-grained objects, such as memory regions and threads, inside the same process. 

    \textbf{Efficiency.} The majority of BCI applications rely on devices that demand solutions with a small memory footprint and slowdown. Therefore, it demands for a lightweight framework suitable for resource-constrained devices.

    \textbf{Practicality.} An easy-adoptable solution should not depend on a specific programming language, and it should be compatible with legacy OSs, libraries, and ML frameworks. Additionally, it should be enabled with only minor programming effort.
\begin{figure}[t]
\centering
\includegraphics[width=.9\columnwidth]{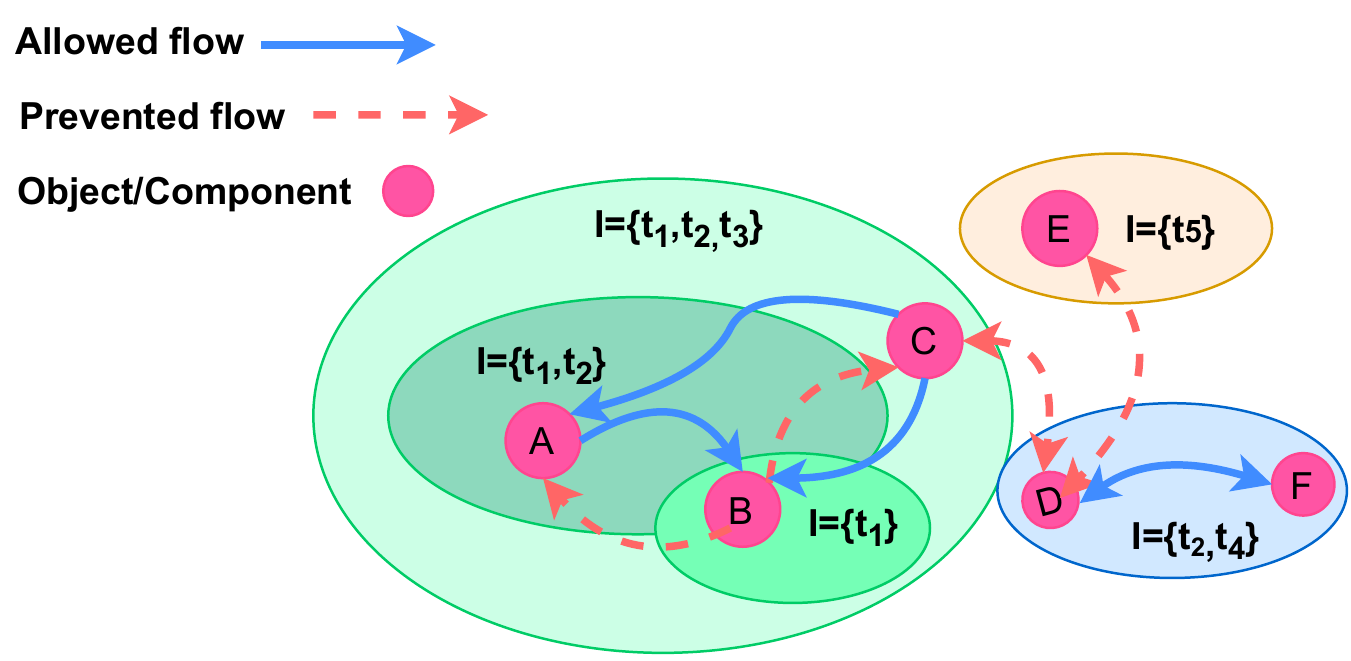}
\vspace{-0.10in}
\caption{\small{\tool: checking integrity information flows.}}
\label{fig:iflow}
\squeezeup
\end{figure}
\subsection{Information flow tracking in \tool}\label{sec:tmodel}
To systematically model the different attack vectors (\S\ref{sec:state}), there is a need for a mechanism that {\em simultaneously} resolves two fundamental issues: lack of extensibility and coarse-granularity. 
To this purpose, our framework provides a set of clear rules for arbitrary types of objects (see Table~\ref{tab:detect}). Then it models each system resource (e.g., memory, thread, file) as an \textbf{\em object} to ensure fine-grained policies. \tool currently supports key objects widely involved in most attack vectors: threads, processes, address spaces, files, pipes, and sockets.
This way, our model allows investigating attacks in a modular manner, where each part of an BCI application (or compartment) can be modeled as a set of system objects. Then our API allows defining per-compartment security policies (e.g., integrity or confidentiality) over the associated objects. Given such a policy, \tool' main goal is to trace and restrict any policy violation.

Our model is inspired by the core concepts of decentralized-IFC~\cite{myers1997decentralized,krohn2007information}, such that every execution entity, like threads or processes, could define and check information flows within the system object itself. \tool uses \textbf{\em tags} and \textbf{\em labels} to trace the data as it flows through the system. To express confidentiality and integrity policies over objects (e.g., file), it assigns them with secrecy or integrity tags. Labels are sets of tags, and each tracked thread $t$ could have two labels: (i) a secrecy label $S_{t}$ to keep all secrecy tags and (ii) an integrity label $I_{t}$ for integrity tags.
If $\delta \in S_{t}$ then the system assumes that thread $t$ has accessed some private data tagged with $\delta$. As such $t$ cannot reveal the seen data or propagate it to another thread that does not have $\delta$ in its secrecy label. To simplify, a secrecy flow from $\alpha$ to $\beta$ is safe when $\{S_{\alpha}\subseteq{S_{\beta}}\}$. Figure~\ref{fig:sflow} illustrates secrecy policy example of safe (e.g., $A$ to $B$) and unsafe (e.g., $C$ to $D$) information flows within a system's components.
Similarly, Figure~\ref{fig:iflow} shows simple integrity-only information flows, where, for example, a flow from $C$ to $A$ and $B$ is allowed since $\{A,B\} \subseteq C$ but $D$ cannot modify $C$ as it does not have a valid label or authorisation ($C \nsubseteq D$). Following this model, Table~\ref{tab:detect} shows simplified policies to tackle our targeted attack vectors.

\section{\tool Implementation}\label{sec:implementation}
Here we describe our prototype of \tool and its components on Linux-ARM platforms.
\subsection{Lib\tool}
\tool provides a new API to enable information flow policies without dealing with the details of the underlying labeling concepts (summarised in Table~\ref{tab:pm}). 
Our user space library supplies seamless tracing with few code changes. After switching to the tracing mode (via \texttt{a\_enable}), any execution thread inside a compartment can assign/remove secrecy or integrity policies over a single or multiple objects (via \texttt{a\_add, a\_remove} or config files). This leads to automatic object initialization and labeling. Developers only need to specify confidentiality or integrity concerns by passing \texttt{SLABEL} or \texttt{ILABEL} flags. As an example, Listing~\ref{tty} shows the API usage for monitoring input injections into the OpenBCI \texttt{tty} ports (e.g., via malicious \texttt{ioctl} calls).

\begin{lstlisting}[caption={\tool: tracing OpenBCI \texttt{tty} port accesses.},captionpos=b,label={tty}]
//*** BCI App side ***//
a_enable();//enable tracing in this compartment
//use SLABEL for secrecy violations
a_add("/dev/tty.OpenBCI-DM00DRM0",file,SLABEL);
//..... rest of the code
//*** attacker side ***//
if __name__ == '__main__':
    port = '/dev/tty.OpenBCI-DN0096XA'
//....initialisation
//Argus detects an access
    board = bci.OpenBCIBoard(port=port, scaled_output=False, log=True)
//..... rest of the code
\end{lstlisting}

\tool also supports tainted memory regions for monitoring accidental or malicious memory operations on sensitive parts of the code. For instance, Listing~\ref{mmio} shows a slight modification in the BrainFlow library to check unauthorised accesses to the physical memory (\texttt{/dev/mem}) used for MMIO (memory-mapped IO) by mapping it to a tainted domain $mobj$ (via \texttt{a\_mmap}).

\begin{table}[htb]
\caption{Lib\tool simplified interface.} 
\vspace{-0.10in}
\centering
\resizebox{.95\columnwidth}{!}{%
\begin{tabular}{|l|l|}
\hline
\textbf{API} & \textbf{Description} \\ \hline
\textbf{New syscalls} & \\ 
a\_enable() \textgreater ret  & enable tracing \\ 
a\_add(obj,type,policy)-\textgreater ret & start tracing object  \\ 
a\_remove(obj,type,policy)-\textgreater{}ret & stop tracing obj \\ 

a\_cleanup(cid)-\textgreater{}ret      & cleanup the tracing                       \\ 
a\_clone(\&fn,policy,...)-\textgreater{}tid     & create a new labeled thread                                           \\ 
a\_execv(mobj,bin...)    & trace binary execution                                          \\ 
a\_wait(\&policy)    & wait for labeled threads                                            \\ 
a\_create(hw\_mode)\textgreater{} mobj   &   create a memory object                                              \\ 

  \hline
  
\textbf{Tainting memory regions} & \\ 
a\_malloc(mobj,size)      &   allocation from mobj       \\ 
a\_free(mobj,size)      &   deallocation from mobj       \\ 
a\_mprotect(mobj,...)   &        change permissions of mobj      \\ 
a\_mmap/munmap(mobj,...)   &    change layout of mobj           \\ 
memcpy,memcmp,memset,etc & other memory operations \\
\hline
  
\textbf{Modified syscalls} & \\ 

open(*pathname, SLABEL|ILABEL,...) &      create labeled file\\

socket(domain, SLABEL|ILABEL,...) &      create labeled socket\\

pipe(pipefd, SLABEL|ILABEL,...) &      create labeled pipe\\


                         \hline
\end{tabular}}
\label{tab:pm}
\end{table}

\begin{lstlisting}[caption={\tool: tracing MMIO accesses in BrainFlow.},captionpos=b,label={mmio}]
int mmio_open(mmio_t *mmio, uintptr_t base, size_t size) {
    return mmio_open_advanced(mmio, base, size, "/dev/mem");}

int mmio_open_advanced(mmio_t *mmio, uintptr_t base, size_t size, const char *path) {
//....initialisation//
 int mobj= a_create(SLABEL);//tainted region
/* Map memory to a tainted memory domain */
if ((mmio->ptr = a_mmap(mobj, mmio->aligned_size, PROT_READ | PROT_WRITE, MAP_SHARED, fd, mmio->aligned_base)) == MAP_FAILED) 
    //... the rest of code...//
    return 0;}
\end{lstlisting}

\subsection{\tool TPMS}

 \kmod implements the described labeling mechanism (see \S~\ref{sec:tmodel}), a set of clear rules and few security hooks to check safe information flows (e.g., \texttt{check\_flow\_allowed}). These security hooks are injected into the rest of the kernel to govern the information flow control. Furthermore, \kmod initialises the required data structures for kernel objects, such as the label registry that caches labels and capability lists. We implemented a hash table-based registry to make labeling data structures more efficient (store/set/get/remove). We set the default maximum label count to $1024$ (configurable). 

\textbf{Tracing threads and processes.}
\kmod stores labels and metadata required for checking dataflows inside each thread's or process' \texttt{cred} structure. When a tainted thread accesses any system object, the system checks its label (\texttt{cred->label}) alongside the object's label to detect information flow violations. 
In addition, to integrate this functionality to the user space, we provide a \tool-assisted multithreading API on top of \texttt{pthread} library for monitoring threads (e.g., to investigate concurrency attacks).

\textbf{Tracing memory regions}.
To efficiently monitor virtual memory objects, \tool implements a small virtual memory abstraction on top of the Linux kernel  ($\approx 3$K LoC) memory management to track and label contiguous virtual memory regions. It uses hardware mechanisms such as ASID (address space identifier) tags and memory domains for fast domain switching and permission changes to reduce the number of TLB flushes. Since BCI applications do not require a large number of tainted memory regions, the available memory domains (16 in ARM-v7) are sufficient, though there are several approaches to scale memory domain usage~\cite{tarkhani2020mu}.
The virtual memory regions are implemented as a per-thread list of contiguous segmented memory blocks, where we store metadata containing their range, label, and permission.

Additionally, we inserted \tool' security hooks in \texttt{mmap.c} to monitor virtual address object mappings by regular applications and separately implement \texttt{a\_mmap/mumap} operations inside the \kmod for tainting memory regions (for instance, IFC checks on MMIO operations).
These security hooks are inserted in critical places to detect changes of the memory layout or permissions of a tainted region; specifically, within memory partitioning structures (e.g., \texttt{\_pgtable\_alloc} and \texttt{init\_pte} in \texttt{mmu.c}).

\textbf{Tracing files, sockets, and pipes.} Our kernel module extends the VFS (virtual file system) layer to monitor security policies within \texttt{inode}, \texttt{file}, and VFS address space operations.  
These objects are used  for operations on unopened files and file handles (including sockets and pipes).
Most \texttt{inode} operations (e.g., \texttt{create, link, mknod}) require a lookup path to find related \texttt{inodes} and \texttt{dcaches}. Hence, we provided a proxy layer on top of the kernel \texttt{namei} to locate \texttt{file/inode\_permission} security hooks and check unauthorised operations like \texttt{read/write/stat/seek} when a tainted object is involved. 
A malicious thread may also try to map a labeled file to an address space object via \texttt{writepage}. \tool makes sure that tainted files are only mapped to tainted memory regions with the right labels via \texttt{a\_mmap}. In addition, it emulates \texttt{open,socket,pipe} system calls to support two new flags (\texttt{SLABEL} and \texttt{ILABEL}), so any thread could easily create a labelled file (e.g. \texttt{O\_CREAT} | \texttt{SLABEL}).
Similarly, \tool inserts hooks for checking security-sensitive \texttt{socket} operations like \texttt{create}, \texttt{listen}, \texttt{connect}, \texttt{sendmsg}, and \texttt{recvmsg}.  The hooks are mostly placed immediately after the lookup process (e.g., \texttt{sockfd\_lookup\_light}). 

\section{Evaluation}
\label{sec:eval}

\textbf{Goals.}
We first describe the implementation details and results of our six attack vectors. Then we evaluate our framework to answer the following questions: \nom{1} Is \tool effective at improving security and privacy on real-world BCI platforms? How does it perform on different attack vectors (AV)? \nom{2} What is the performance overhead? Is it suitable for resource-constrained devices? \nom{3} How practical is \tool from the developers' perspective (e.g., programmability), compared to other potential approaches?

\textbf{Setup.} We prototype Argus on the Raspberry Pi 3 (rpi) Model B~\cite{rpi3} with a 1.2 GHz 64-bit quad-core ARM Cortex-A53 processor with 32KB L1 and 512KB L2 cache memory, running a 32-bit unmodified Linux kernel version $4.19.42$ and glibc $v2.28$ as the baseline.
We adopt two sets of microbenchmarks, LMbench 3.0~\cite{mcvoy1996lmbench} and a custom benchmark to measure overhead on different functionalities. Additionally, we use the following standard BCI boards and their unmodified libraries.

\textbf{OpenBCI.} The OpenBCI Cyton Biosensing Board is an 8-channel neural interface with a 32-bit processor.  It implements the PIC32MX250F128B microcontroller and relies on the chipKIT bootloader and OpenBCI firmware. Data is sampled at 250Hz on each channel. The board communicates wirelessly to any mobile device or tablet compatible with Bluetooth low energy (BLE). We use version $3.1.2$ of OpenBCI cyton libraries, its SDK ($v3.0.0$), and BrainFlow (v4.8.2).

\textbf{NeuroSky.} The NeuroSky MindWave Mobile 2-EEG (MWM2) brainwave headset uses the ThinkGear ASIC chip (TGAT1 ASIC) with a static headset MAC address (ID) and has a single EEG channel. It relies on BLE for communications. We use NeuroSky Android developer tools and libraries (v4.2). 

\textbf{Muse2.} This headset is widely adopted for mental health evaluation. It is equipped with seven sensors, four of which are EEG channels. The headset uses a PIC24 microcontroller and RN42 Bluetooth chip for BLE-based communication. 
Many Muse-based applications also rely on BrainFlow for obtaining and processing EEG data.

\subsection{Attack PoCs: threats from/to the host}\label{sec:hostatt}
We implemented multiple PoCs of AV1, AV2, and AV3 on Muse, NeuroSky, and OpenBCI platforms.

\textbf{Sniffing.}
We implemented and tested several BLE sniffers on the three devices and analyzed their transmitted packets (e.g., via Wireshark and nRF52840 Dongle). 
To reflect the diversities between these platforms in terms of software dependencies and transmission protocols, we built different attack PoCs for each device. For instance, MWM2 uses BLE(GATT) to communicate with iOS devices without relying on encryption to protect its connections. The headsets transmit several types of data, including signal quality (how well the device is picking up brainwaves) and raw EEG data. The GATT service used is not standard, and details are not documented or open-source. Hence, we discovered most details by sniffing the communication between MWM2 and an iPad. We manually analysed and detected GATT characteristics used in the device notifications (based on the frequency of the notifications) to further discover which ones were attention or meditation values (for instance, \texttt{0x001C} in Listing~\ref{nsky} for attention signals).
We then implemented a sniffing plugin summarised in Listing~\ref{nsky}, for displaying different signal values directly through Wireshark. Muse2 and OpenBCI rely on similar BLE connections but with different types of packets. Nevertheless, we could extract all important transmitted data in plain text on these platforms too using a similar approach.\\

\textbf{MITM.}
We successfully launched MITM on all devices 
to record and store all transmitting communications to an external device, impersonate the BCI headset, and alter the transmitted information. We used two Raspberry Pis and existing tools for device emulation and interception (e.g., via GATTacker with Node.js v8.9.0~\cite{jasek2016gattacking}).
One rpi acts as a BLE central device and connects to the headset; the other rpi acts as a BLE peripheral device and transmits advertising packets pretending to be the headset. For Neurosky and Muse, we did not need to spoof the MAC address of the headset by the peripheral Pi, since their apps does not check the MAC addresses of the devices it connects to. The user's device will then connect to the peripheral Pi, believing it to be the headset. At this stage, we could intercept and modify all the data sent between the headset and the host device since the data passes through the two rpis. Note that these PoCs are only a subset of all possible MITM attacks that can happen on current BCI applications.

\begin{lstlisting}[caption={Pseudocode of sniffing plugin for  NeuroSky.},captionpos=b,label={nsky}]
&\color{teal}-- our pseudo protocol&
neurosky_proto = Proto("neurosky","NeuroSky Postdissector")
attention_F = ProtoField.string("neurosky.attention","Attention")
&\color{teal}-- similar to other metrics like meditation&
&\color{teal}-- add the field to the protocol&
neurosky_proto.fields = {attention_F, meditation_F, signal_F}
&\color{teal}-- create a function to "postdissect" each frame&
function neurosky_proto.dissector(buffer,pinfo,tree)
if btatt_handle_f() and btatt_value_f() and btatt_opcode_f() then
    &\color{teal}-- obtain the current values of the protocol fields&
    local btatt_value = tostring(btatt_value_f())
    if btatt_handle == "0x0000001c" and btatt_opcode == "0x0000001b" and string.sub(btatt_value, 7, 8) == "ea" then
        local subtree = tree:add(neurosky_proto,"NeuroSky Protocol Data")
        subtree:add(attention_F,tonumber(string.sub(btatt_value, 25, 26), 16))
    end
&\color{teal}-- register our protocol&
register_postdissector(neurosky_proto)

\end{lstlisting}

\textbf{Inadequate access control \& privilege escalation.} We implemented several attack PoCs to discover that all the considered BCI use cases and devices suffer from insufficient access control and privilege management. For example, NeuroSky's ThinkGear Connector (TGC) runs as a background process on the host device and is responsible for directing headset data from the serial port to an open network socket. For easy integration, it is designed to communicate with any framework (via sockets), such as Adobe Flash. Hence, any malicious process running on a user's computer can ask TGC for brainwave data with no specific permission. 
To explore further, we launched a few processes for connecting to the headset through TGC, getting all private data (e.g., attention and meditation data), and then transmitting this data to a remote device over the internet. 
When TGC is already sending data to a process, it will not notify the user if and when another application connects to it. The only way the user could find out that multiple applications are receiving their private data is to attempt to monitor active processes manually.
Additionally, we could impersonate the TGC by simply listening on its unprotected TCP port $13854$ and sending brainwave data to other untrusted processes.

As another PoC, we examined eegID~\cite{eegid}, an Android application that visualises and records data from a NeuroSky headset. It stores the collected EEG and GPS data on the mobile phone's storage (in eegIDRecord.csv file). 
Hence, any process with ``Storage'' permission can read this data, which means potentially any other app could gain access to the location and brainwave data of the user. Even if originally the malicious app does not have permission for recording GPS data, it can now access it through the vulnerable BCI application.
We implemented and confirmed a similar attack by compromising Muse's Mind Monitor, an app that lets users view and record data from a Muse headset (unlike eegID, Mind Monitor does not store GPS data). Similarly, this app does not isolate its storage. 

On OpenBCI, we found similar vulnerabilities ranging from memory corruption (mostly due to unsafe C API, such as \texttt{memcpy}, \texttt{strcpy}, or \texttt{strlen}), race conditions (e.g., via TOCTOU or Python subprocesses), and inadequate isolation or access control (e.g., unprotected storage/SDCard, unencrypted BLE communication, or incorrect process permission setup) as summarised in Table~\ref{tab:bflow}. Additionally, the lack of privilege separation within different modules, allowed us to execute arbitrary code on the host with root privilege by exploiting the BrainFlow vulnerabilities and gaining ``root access'' through the OpenBCI user interface.

\subsection{Attack PoCs: ML adversarial threats}\label{sec:mlatt}
To investigate the effectiveness of Argus in detecting adversarial attacks to BCI deep learning models, we prototype the Avs mentioned in \S~\ref{sec:adversarial_ml} on six dataset-model combinations. The aim is to showcase how Argus can inform the deep learning developer about the model vulnerabilities and how the BCI model behaves under attack.
\begin{figure}
    \vspace{-0.20in}
    \centering
    \includegraphics[width=0.5\textwidth]{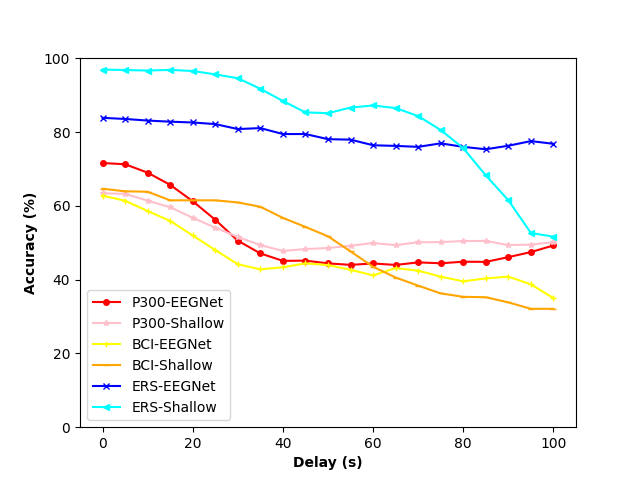}
    \vspace{-0.35in}
    \caption{Peripheral injection adversarial black-box attack on the 3 dataset-model combinations.}
    \label{fig:delay_attack}
    \vspace{-0.25in}
\end{figure}

\textbf{Datasets.}
The BCI platforms considered in this work (and in general) do not release the collected data, therefore we evaluate the adversarial attacks on three publicly-available EEG datasets. We consider data including both univariate (as in NeuroSky) and multivariate (as in Muse and OpenBCI) timeseries of EEG signals (depending on the number of electrodes) similar to the use cases presented by the three BCI platforms we study.

\textit{P300 Evoked Potential~\cite{P300}} represents P300 evoked potentials recorded with the BCI2000~\cite{schalk2004bci2000} interface. Each subject was presented with a 6x6 character matrix with randomly flashed elements. The participant was asked to focus on the target character, and count the flashing repetitions of the row and column containing the target character. EEG signals are collected by a 64-electrode scalp, sampled at 240 Hz, and bandpass filtered between 0.1 Hz and 60 Hz.

\textit{Epileptic Seizure Recognition (ESR)~\cite{andrzejak2001indications}}, extracted from 500 patients, contains single-channel EEG signals of length 178. Each signal falls into one of 5 classes: normal patient eyes open, normal patient eyes closed, tumor patient healthy area, tumor patient tumor area, and epileptic patient seizure. No further preprocessing is applied.

\textit{BCI-IV-2a~\cite{tangermann2012review}} contains recordings of four motor imaginary tasks (imagination
of the movement of the left hand, right hand, both feet, and tongue) performed by nine subjects across 22 EEG channels. The signals were sampled with a 250 Hz  and bandpass filtered between 0.5 Hz and 100 Hz. Additionally, a notch filter (50 Hz) is enabled.

\textbf{Deep learning architectures.}
Each of the datasets is paired with two state-of-the-art EEG classification deep learning architectures described as follows.

\textit{ShallowCNN~\cite{schirrmeister2017deep}} is a network specifically designed for BCI applications. The first two layers consist of temporal and spatial convolutions which correspond to transformations made during the bandpass and spatial filtering stages of filter bank common spatial patterns (FBCSP), a traditional approach to feature extraction widely used across multiple BCI applications. The output feature map is then fed to a non-linearity, pooling layer and a logarithmic activation. Batch normalization and dropout (with rate 0.5) are added as regularizers.

\textit{EEGNet~\cite{lawhern2018eegnet}} is a CNN-based EEG classifier. In the first block, a 2D convolutional layer outputs feature maps containing the EEG signals at different frequencies. Then the depthwise convolution is used to learn a frequency-specific spatial filter. In the second block of the network, a separable convolution is applied, where a depthwise convolution is followed by a pointwise convolution. This architecture allows for parameter reduction and decoupling of relationships across feature maps. Additionally, batch normalization, dropout (with rate 0.5) and ELU non-linearity are used.

\textbf{Adversarial attack PoCs.} The deep learning models are implemented in PyTorch~\cite{paszke2019pytorch} and quantized (post training) to use 8-bit weights and activations to run on the rpi. The training procedure does not include any adversarial training or hyper-parameter tuning to increase robustness. We implement AV4, AV5 and AV6 in PyTorch, too, although these attacks are ML framework independent. For AV4 (FGSM) and AV5 (PGD), we evaluate perturbation strengths in $\epsilon = \{0.01,0.02,0.03,0.04,0.05,0.6,0.07,0.5,1\}$. PGD is an iterative attack and in our experiments we use iterations $ n \in$ \{2,3,4,5,6,7,8,20,40\} (attack applied in iterative manner as explained in \S~\ref{sec:adversarial_ml}). The peripheral injection black-box attack (AV6) perturbs the input to introduce a delay corresponding to $\tau/f$ where frequency $f$ is 240Hz, 178Hz, and 250Hz for P300, ESR and BCI-IV-2a, respectively. The adversarial perturbations are applied to the whole test set and all the metrics represent the overall accuracy when an adversarial attack targets each sample.
\begin{figure}[t]
    \centering
    \includegraphics[width=0.5\textwidth]{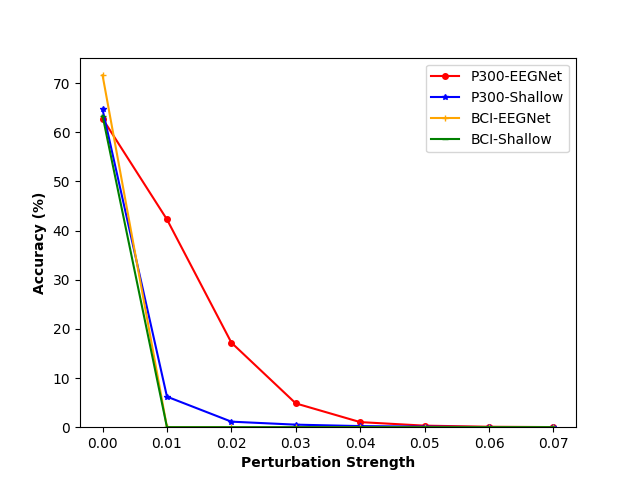}
    \caption{FGSM (AV4) adversarial attack at different $\epsilon$ values.}
    \label{fig:fgsm_attack}
\end{figure}

Figure~\ref{fig:delay_attack} shows how the accuracy of the EEG models degrades at the presence of AV6 at different delay values. Each dataset and model have a different vulnerability pattern depending on the delay value suggesting that specific mitigation is needed for each of them. Although in some scenarios ShallowCNN presents higher initial accuracy on clean samples (especially for P300 and ERS), it shows higher adversarial vulnerability at higher delay values compared to EEGNet. Figure~\ref{fig:fgsm_attack} and~\ref{fig:pgd_attack}, instead, show the attack vectors AV4 and AV5, respectively. For FGSM, all models show a drastic decrease in accuracy even at low values of adversarial strength as represented by a small $\epsilon$. For PGD, too, the models drop to zero accuracy faster as here the attack was applied a higher strength $\epsilon=0.5$ and several iterations for a bigger adversarial impact were performed. 
It is noticeable that the latter attack converges fast enough (few iterations needed) to be cost-efficient and widely attack at a high frequency. Given the nearly-immediate impact on the prediction accuracy, these attacks can cause real-time damages (for instance, in the case of BCI-assisted wheelchair or smart car).
To conclude, our PoCs show that EEG-based BCI deep learning models are extremely vulnerable to a variety of adversarial AVs.

\subsection{\tool Microbenchmarks}
\textbf{System-wide effects.} 
We enabled \tool on Linux with lightweight kernel module ($\approx{8}$K LoC) and userspace library ($\approx{2}$K LoC).
To evaluate \tool' effect on theoverall performance of Linux sub systems, we used LMBench (Figure~\ref{fig:lmbench}). Our results show $\approx 16\%$ latency overhead for FS, $\approx 0.6\%$ for networking, $\approx 0.2\%$ for IPC benchmarks, and $0.1\%$ for multi-threading in worst cases.
%
%
It is difficult to directly compare these results with previous information flow tracking systems for edge use cases since they are implemented as programming language extensions or specific to Android ecosystems (e.g., TaintDroid).
\begin{figure}[t!]
    \centering
    \includegraphics[width=0.5\textwidth]{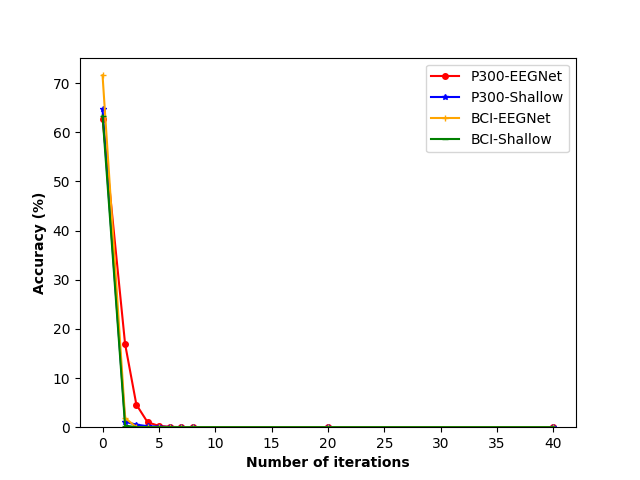}
    \caption{PGD (AV5, $\epsilon=0.5$) attack for different iterations.}
    \label{fig:pgd_attack}
\end{figure}
 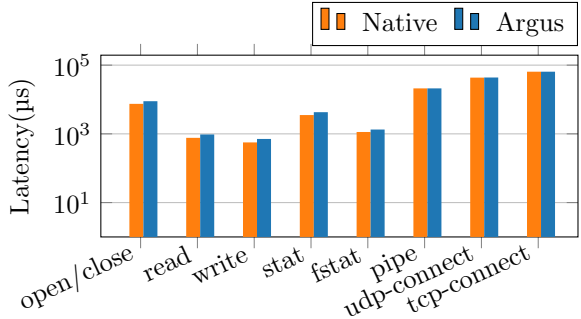
\begin{figure}[htb]
 \begin{tikzpicture}
    \begin{axis}[
         width=0.48\textwidth,
        height=4cm,
        bar width=5.3pt,
   symbolic x coords={open/close,read,write,stat,fstat, pipe, udp-connect, tcp-connect},        
        xlabel={},
        xtick=data,
        x tick label style={rotate=25,anchor=east},
        ymin=1,
        ybar=0pt,
        ymajorgrids=true,
        ylabel={{Latency(\textmu s)}},
        ymode=log,
        legend cell align=left,
  legend style={
            at={(1,1.05)},
            anchor=south east,
            legend columns=4,
            column sep=1ex
        },
    ]
        \begin{scope}[
            xshift={0.1*\pgfplotbarwidth},
            draw=none,
        ]

        \addplot [draw=none, fill=torange] coordinates {
        (read,765.7)
        (write,563.4)
         (stat,3524.7) 
        (fstat,1133.5)
        (open/close,7465.1)
        (pipe,21042)
         (udp-connect,43293)
        (tcp-connect,64396)};
         \addplot [draw=none, fill=b1] coordinates {
        (read,959.5)
        (write,711.1)
         (stat,4272.4) 
        (fstat,1336.1)
        (open/close,8911.9)
        (pipe,21053) 
        (udp-connect,43639)
        (tcp-connect,64401)};

        \end{scope}

\legend{Native, Argus}
    \end{axis}
\end{tikzpicture}

\caption{Overall \tool overhead on Lmbench. The term \textit{native} is used for unmodified applications or benchmarks running on an unmodified Linux kernel.}
\label{fig:lmbench}
 \end{figure}

\textbf{Lib\tool overhead.}
For enabling dataflow tracking, Lib\tool adds an $\approx 33 \mu s$  latency overhead. Table~\ref{tab:api} and ~\ref{tab:overhead3} summarise the performance overhead of using Lib\tool' APIs. 
Unlike previous IFC systems, \tool enforces 
intra-application policies via tainted memory regions instead of relying on specific programming language (e.g., via labeling Java classes).
Creating a \tool's tainted memory (via \texttt{a\_create}) and mapping or unmapping memory pages to it (\texttt{a\_mmap, a\_munmap}) adds a $4.8\%$ (stddev $0.17\%$) overhead on average. This is usually a one-time operation at the initial phase of an application. 
Lib\tool memory management (\texttt{a\_malloc/free}) is $0.03\%$ faster than unmodified malloc/free in our microbenchamrks due to using fixed size regions and optimisation using ARM memory domains. 
%
Increasing size of labels only effects (linearly) the performance of operations on 
that labeled object, without additional cost for untainted objects. 

\subsection{Real-world use cases}

There are different ways to use Lib\tool. On one hand, \emph{BCI framework/SDK} developers can integrate it into their codebase via small modifications. On the other hand, \emph{BCI application} developers that need to rely on insecure and closed-source BCI third-party libraries, can utilize Lib\tool to track information flows within all these untrusted modules alongside their own applications' sensitive dataflows.
We successfully detect and avoid the six AVs we discussed earlier (\S\ref{sec:state}) by employing \tool with reasonable effort and overhead. Further, we connect different BCI devices to rpi3 to evaluate all of them on the same baseline, running on the same \tool-enabled Linux kernel. The following summarises our specified information flow policies, modifications, and results on each BCI platform:

 \begin{table}[t]
 \caption{Cost of initialing \tool (average $5000$ run).}
 \vspace{-0.10in}
\centering
\LARGE
 \resizebox{\columnwidth}{!}{%

    \begin{tabular}{|c|ccccc|}
    \hline
    Lib\tool API        & a\_enable & a\_cleanup & a\_add & a\_remove & a\_create \\ \hline
    Latency (\textmu s) & 30.1      & 35.8       & 32.3   & 32.82     & 32.9      \\ \hline
    \end{tabular}}
\vspace{-0.25in}
\label{tab:api}
\end{table}

 \begin{table}[t]
 \caption{Overhead of labeled multi-threading and memory operations compared to native (average $20000$ run). We compared \texttt{a\_clone} cost with \texttt{pthread} (glibc 2.28).}
\centering
 \LARGE
 \resizebox{\columnwidth}{!}{%

    \begin{tabular}{|c|ccccc|}
    \hline
    Lib\tool API        & a\_mmap & a\_malloc/free & a\_clone & a\_wait & a\_execv \\ \hline
    Overhead (\%) & +4.8      & -0.03        & +2.5   & +1.8     & +1.1      \\ \hline
    \end{tabular}}
\label{tab:overhead3}
\end{table}

\begin{table}[t]
    \caption{Overhead of integrating and using Lib\tool.}
    \vspace{-0.10in}
 \centering
 \resizebox{\columnwidth}{!}{%
    \begin{tabular}{|l|l|l|}
    \hline
    BCI platform & Modified/added LoC & Slowdown \\ \hline
    OpenBCI       & 0.02\%   & 14.2\%     \\
        Muse       & 0.01\%   & 12.1\%     \\ 

    NeuroSky    & 0.01\%   & 12.7\%     \\ \hline
    \end{tabular}}
    \vspace{-0.20in}
    \label{tab:libuse}
\end{table}

\textbf{OpenBCI-based platforms.} Instead of modifying separate OpenBCI applications, we opted for modifying the underlying open-source library, Brainflow, which is used by most applications.
After enabling \tool in the initialization phase (via \texttt{a\_enable}), we modify several parts of BrainFlow for tainting and tracking various objects. This includes modifying the modules that handle connections to the BCI headset to enable labelling the serial/BLE ports, sockets, and worker threads. Additionally, we label any file that contains the headset information, metadata, and logs. This required small modifications for example, in \texttt{board\_shim} , \texttt{brainflow\_get\_data}, and \texttt{brainflow\_filter}. To avoid data leaks at runtime (e.g., via memory corruption attacks), we assigned a tainted memory region for the headset information and another one for mapping the log file. Similarly, we labeled the ML model and mapped it to a separate tainted memory region. Table~\ref{tab:libuse} summarises the overhead introduced by our changes.

\textbf{Muse \& NeuroSky-based platforms.}
Many Muse applications depend on BrainFlow too, therefore the same modifications as OpenBCI are required. Additionally, some Muse applications use Muse-lsl library\footnote{https://github.com/alexandrebarachant/muse-lsl} for streaming, visualizing, and recording EEG data. We integrated Lib\tool to different parts of muse-lsl such as \texttt{muse} backend, \texttt{record}, and \texttt{stream} to taint all sensitive threads, files, and communication ports with small modifications as shown in Table~\ref{tab:libuse}. 
NeuroSky does not officially provide Linux-based libraries, so most of its Linux-based applications use python-mindwave\footnote{https://github.com/akloster/python-mindwave} library. Therefore, we integrated Lib\tool to this library by applying modifications similar to the muse-lsl ones.

\textbf{Utilising Lib\tool for ML adversarial attacks.}
For all the BCI uses cases we rely on the deep learning framework described in \S~\ref{sec:mlatt}.
Through \tool, we could detect the aforementioned adversarial attacks by carefully modeling and mapping the attack vectors into the underlying system objects (e.g., threads, files, memory regions, device peripherals) and then monitoring the objects. For instance, we tracked information flows during white-box attacks (AV4 and AV5) by tainting the BCI process and model file $f_{model}$ in the file system as well as the memory blocks (e.g., $M_b\{m_i,\dots,m_j\}$ in Table~\ref{tab:detect}) assigned to the model (e.g., upon the \texttt{mmap}). This enables the system to detect any information flow policy violation ($obj \not\longrightarrow P_{\otimes}$/$P_{\otimes} \not\longrightarrow obj$ in Table~\ref{tab:detect}) which allows the attack to get access to the deep learning model in the first place. Furthermore, we know that the attacker needs to re-inject the perturbed signal therefore we could track this injection by monitoring the BCI device ports (e.g., $S_b\{port_{in}\}$ in Table~\ref{tab:detect}). These are just a set of examples on how we used \tool for tracing ML adversarial attacks, which can be easily extended to future unseen attacks.

\section{Conclusion}
\label{sec:conclusion}

We presented a two-fold contribution for enhancing security and privacy on wearable BCI platforms. 
We prototyped six major attack vectors including system-side and adversarial ML attacks on real-world BCI use cases. 
We then introduced \tool, the first information flow control system for wearable BCIs.
%
We also discovered a sheer number of vulnerabilities (> 300), highlighting the subsequent importance of techniques like \tool.
We believe that the simplicity of our framework, combined with its extensibility and fine-granularity features, position \tool as the first system for future work on security and privacy of wearable BCIs.


\balance
{\footnotesize \bibliographystyle{acm}
\bibliography{main}}



\end{document}